\documentclass[
  ,final            
  ]
  {aipproc}

\layoutstyle{6x9}

\newcommand{\comment}[1]{{}}

\begin{document}

\title[Global Analyses of Nuclear PDFs]{Global Analyses of Nuclear PDFs\footnote{Talk given at DIS 2005, Madison, USA, April 27-May 1, 2005.}}

\classification{24.85.+p}
\keywords      {Global analysis, nuclear PDF, shadowing}

\author{V.J.~Kolhinen}{
  address={Helsinki Institute of Physics, P.O. Box 64, 
  FIN-00014 University of Helsinki, Finland \\
  Department of Physics, P.O.Box 35 (YFL), 
  FIN-40014 University of Jyv{\"a}skyl{\"a}, Finland}
}

\begin{abstract}
  A brief overview of the global DGLAP analyses of the nuclear parton
  distribution functions is given. Although all the current global
  nPDF sets describe $R_{F_2}^A(x,Q^2)$ well in the large-$x$ region
  where the data exist, variations between their parton distributions
  can be substantial.
\end{abstract}

\maketitle



Ever since it was observed about two decades ago that parton
distribution functions (PDFs) of nuclei differ from those in the free
proton, $f_i^A(x,Q^2) \ne f_i(x,Q^2)$, several analyses of the nuclear
effects have been presented. Similarly to the case of the free proton
PDFs, the nuclear PDFs (nPDFs) at an initial scale provide the
nonperturbative input for the perturbative QCD analysis. Once they are
known the Dokshitzer-Gribov-Lipatov-Altarelli-Parisi (DGLAP) evolution
equations predict their behaviour at larger scales.

Although several DGLAP analyses
\cite{Qiu:1986wh,Frankfurt:1990xz,Eskola:1992zb,Indumathi:1996pb,
Indumathi:1996ky,Frankfurt:2003zd} exist, I will concentrate here only
on the recent global ones where the initial distributions are based on
the fit to the data, not on a model. Only three such analyses and
their reanalyses currently exist, namely the ones by us, Eskola {\it
et al.} (usually called as {\it EKS98}
\cite{Eskola:1998iy,Eskola:1998df}), Hirai {\it et al.} ({\it HKM}
\cite{Hirai:2001np} and {\it HKN} \cite{Hirai:2004wq}) and by
de~Florian and Sassot ({\it nDS} \cite{deFlorian:2003qf}). Along with
these I will also present preliminary results of the reanalysis of our
nPDFs ({\it EKS05} \cite{EKS05}).

The analyses of the nPDFs are performed much in the same way as those
of the free PDFs. However, lack of data especially in the small-$x$
region has kept the nPDF analyses less constrained than the free PDF
ones. For example, whereas the recent PDF analyses have been performed
in next-to-leading order (NLO) and some are currently being calculated
in NNLO, only one global nPDF analysis, by de Florian and Sassot
(nDS), is currently calculated in NLO.

Since the initial states of the global nPDF sets are based on the data
they naturally describe well the structure function $F_2^A(x,Q^2)$ at
large $x$ where the most of the data lie. However, their nuclear
modifications for the different parton flavours can vary greatly in
some regions.
 
Nuclear effects are commonly defined through a ratio
\begin{equation}
  R_i^A(x,Q^2)=\frac{f_i^A(x,Q^2)}{f_i(x,Q^2)},
\end{equation}
where $f_i$ stands for parton distribution function for a parton type,
$i=u, \bar{u}, d, \bar{d}, \ldots, g$. However, due to lack of data
one usually defines only some 3-5 ratios for the initial
distributions: for valence ($u$ and $d$ together (EKS98), or
separately (HKM,HKN,nDS)), sea ($\bar{u}$ and $\bar{d}$ together
(EKS98,HKM,HKN), or separately (nDS)) and gluon.  These ratios are
usually given for a bound proton.

The $R_{F_2}^A$ ratios are fairly well constrained in large-$x$ region
by the data from lepton-nucleus deep inelastic scattering (DIS). As
valence quarks dominate the $F_2$ in this region, they also become
well determined there.  At mid-$x$ the Drell-Yan (DY) dilepton data
constrain the sea quark distributions together with the DIS data. For
gluon distribution the only data constraints arise from the $\log Q^2$
slopes of the NMC data for $F_2^{\rm Sn}/F_2^{\rm C}$
\cite{Arneodo:1996ru} as shown in Refs.
\cite{Eskola:1998iy,Prytz:1993vr}:
\begin{equation}
  \frac{\partial R_{F_2}^A(x,Q^2)}{\partial \log Q^2}
    \approx
    \frac{10\alpha_s}{27\pi}\frac{xg(2x,Q^2)}{F_2^D(x,Q^2)}
    \biggl\{R_G^A(2x,Q^2)-R_{F_2}^A(x,Q^2)\biggr\}.
\end{equation}
In addition to the data further constraints for the fits arise from
the momentum, charge and baryon number conservation. Let us next take
a look at these three analyses individually.\\

{\bf EKS98, EKS05:} In the nPDF analysis by us, Eskola, Kolhinen,
Ruuskanen and Salgado \cite{Eskola:1998iy,Eskola:1998df} the
$R_{F_2}^A(x,Q_0^2)$ distribution was first parametrized piecewise for
each $A$ and fitted to the data. The $R_{F_2}^A$ was then split to
valence and sea part which were constrained by the DY data and baryon
number conservation. Finally, the gluon distribution was constructed
from the $R_{F_2}^A$ fit. The actual fits to the DIS and DY data were
done by eye. However, later calculations with EKS98 prove $\chi^2
\approx$ 390 for 503 data points, though the data set is slightly
different than in the original analysis.

As a continuation for this work, we are currently performing a
reanalysis of the nPDFs, with some more recent data included
\cite{EKS05}. We have now also included a proper $\chi^2$ analysis and
use a Hessian method for the error estimates. Instead of parametrizing
the $R_{F_2}^A$ as in EKS98, we now parametrize directly the nuclear
effects in the initial valence, sea and gluon distributions. Although
the results are still preliminary, they seem to resemble much the
EKS98 ones and giving $\chi^2 \approx 390-400$ for 503 data points.

{\bf HKM, HKN:} In their first analysis Hirai, Kumano and Miyama (HKM)
\cite{Hirai:2001np} use the DIS, but not DY, data from several
experiments.  They composed two different fits, ``quadratic'' and
``cubic'' referring to the polynomial in the fit. These fits are
performed for each $A$ separately. The resulting valence ratios
$R_u^A$ and $R_d^A$ are not given for bound proton but for an average
nucleon in a nucleus.  The calculated $\chi^2$ of the fit is 583.7
(quadratic) and 546.6 (cubic) for 309 data points, or $\chi^2/d.o.f. =
1.93$ and 1.82, respectively. The nuclear effects show small
antishadowing for the valence at small $x$. Sea and gluons are
shadowed in small-$x$ region, but antishadowed at larger values of
$x$. Only valence shows an EMC effect (shadowing) at $x\sim 0.7$.

In the subsequent analysis by Hirai, Kumano and Nagai (HKN)
\cite{Hirai:2004wq}, the DY data have been included along with some
more DIS data. The statistical error analysis is also performed using
a Hessian method. The general form of the fit is similar to the
``cubic'' form in HKM. The resulting distributions fit the small $x$
region better, obviously due to the DY data and improved description
of the sea quarks. Whereas valence and gluon behave much in the same
way as in the HKM, the sea quarks now have a valence-like EMC effect
at $x\sim 0.7$. The resulting $\chi^2 = 1489.8$ for 951 data points.

{\bf nDS:} The global nPDF analysis by de~Florian and Sassot (nDS)
\cite{deFlorian:2003qf} is so far the only one performed in NLO.  In
this analysis PDFs are defined using the convolution method,
  $ f_i^A(x,Q_0^2) 
    = \int_{x}^A \frac{dy}{y} \, w_i(y,A) f_i(\frac{x}{y},Q_0^2), $
which enables evolution in the Mellin space. The advantages of this
approach are that the calculations are faster and more
straightforward, as well as that the $x$ dependence of nPDFs is
strongly correlated to that of free PDFs.  The total $\chi^2$ obtained
is 316.35 for LO and 300.15 for NLO for 420 data points. The main
difference between LO and NLO results are in sea and gluon
distributions at small $x$, small $Q^2$ and large $A$. As the $\chi^2$
values suggest, LO fit describes the data almost equally well as the
NLO one. This fact is reported to arise from the rather restricted
$Q^2$ range of the data as well as the absence of the data strongly
dependent to gluon distribution.

Compared to the other analyses, the largest difference is in the gluon
distribution, which is much less shadowed than e.g. in EKS98.  The
authors have tried another parametrization with stronger gluon
shadowing, but they report the results to be worse. \\


\begin{figure}
  \includegraphics[height=.39\textheight]{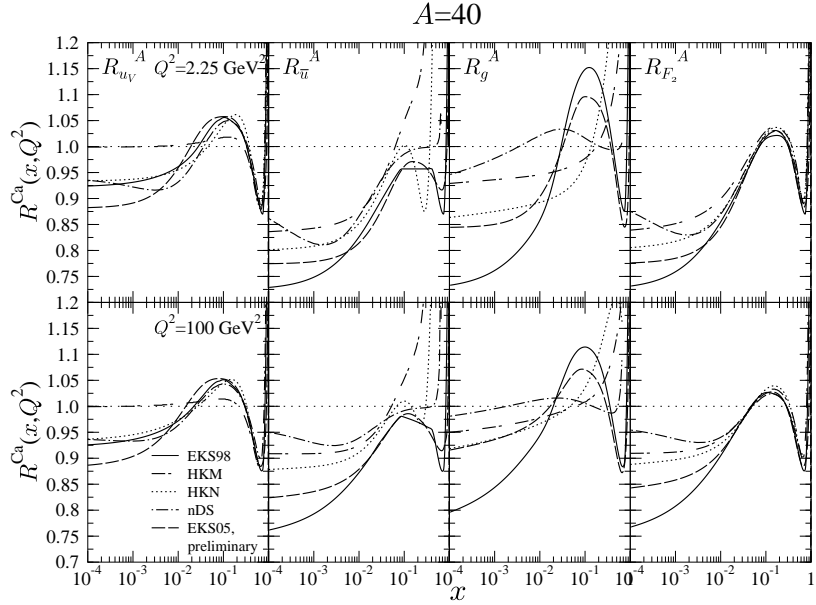}
  \caption{Ratios $R_{u_v}^A(x,Q^2)$, $R_{\bar u}^A(x,Q^2)$,
  $R_g^A(x,Q^2)$ and $R_{F_2}^A(x,Q^2)$ for $Q^2 = 2.25$ GeV$^2$
  (upper panels) and $Q^2=100$ GeV$^2$ (lower panels) for $A=40$ given
  by EKS98 (solid), HKM (double dashed), HKN (dotted) and nDS [NLO]
  (dotted-dashed).  Preliminary EKS05 results are also shown
  (dashed).} \label{RAQ2}
\end{figure}

\begin{figure}
  \includegraphics[height=.39\textheight]{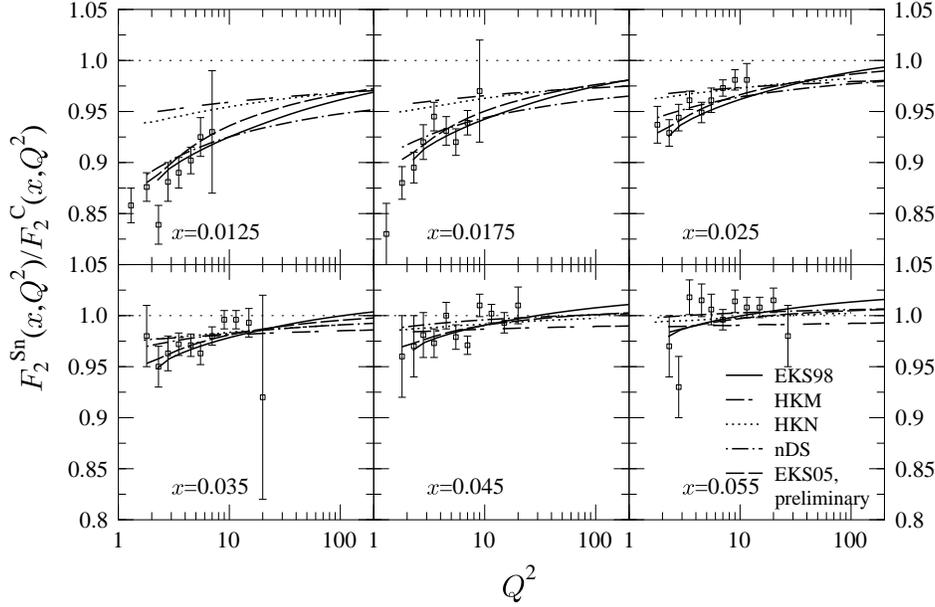}
  \caption{Calculated $F_2^{\rm Sn}/F_2^{\rm C}$ ratios compared to
  the NMC data \cite{Arneodo:1996ru} for a few small $x$ values.}
  \label{SnC}
\end{figure}

Comparison between the nuclear effects of different sets are shown in
Fig. \ref{RAQ2} for $A=40$ and for $Q^2=2.25$ and $100$ GeV$^2$.  As
seen in the figure, $R_{F_2}^A$'s calculated using different nPDF sets
coincide in the large $x$ region. Also valence quarks become well
determined in large $x$. However, in other regions the differences
between the sets can be large.  In order to constrain the fits more
properly, especially at small $x$, more data would be needed.  As
pointed out earlier and shown in Fig. \ref{SnC}, currently only the
$\log Q^2$ slopes of NMC data for $F_2^{\rm Sn}/F_2^{\rm C}$ give
constraints to gluon distributions.  Probes sensitive to nuclear gluon
PDFs, such as the charm production, would thus be crucial for more
accurate analysis.

In future analyses data on structure function $F_3$ could also provide
more information on the $u_v^A$ vs $d_v^A$ ratio and the valence
shadowing.

\paragraph{Acknowledgements}
  This project was funded by the Academy of Finland, projects nos.
  80383 and 206024.



\bibliographystyle{aipproc}   



\end{document}